\documentclass{elsart}

\usepackage{psfig}
\begin{document}
\begin{frontmatter}
\title{Teleportation in the Background of Schwarzschild Space-time}
\author[shao,nao,gs]{Xian-Hui Ge}
\author[shao,nao,itp,email]{You-Gen Shen}
\address[shao]{Shanghai Astronomical Observatory,Chinese Academy of Sciences,
Shanghai 200030, China (post address)}
\address[nao]{National Astronomical Observatories, Chinese Academy of Sciences,
Beijing 100012, China}
\address[gs]{Graduate School of Chinese
Academy of Sciences, Beijing 100039, China}
\address[itp]{Institute of Theoretical Physics, Chinese Academy of Sciences,
Beijing 100080, China}
\thanks[email]{e-mail:ygshen@center.shao.ac.cn}

\begin{abstract}
  \hspace*{7.5mm}Quantum teleportation is investigated between Alice who is
far from the horizon and Bob locates near the event horizon of a
Schwarzschild black hole. The results show that the fidelity of
the teleportation is reduced in this curved space-time. However,
high fidelity can still be reached outside a massive black hole.
\\

\end{abstract}
\end{frontmatter}
\newpage
\hspace*{7.5mm}The new field of quantum information has made rapid
progress in recent years[1,2,3,4]. Relativistic quantum
information theory may become a necessary theory in the near
future, with possible application to quantum teleportation. A
number of authors have studied quantum entanglement in
relativistic frames inertial or not[5-12]. Czachor studied a
version of the electron is paramagnetic resonance experiment with
relativistic particles[10], and Peres et al. demonstrated that the
spin of an electron is not covariant under Lorentz
transformation[7].Moreover, Alsing and Milburn studied the effect
of Lorentz transformation on maximally spin-entangled Bell states
in momentum eigenstates and Gingrich and Adami derived the general
transformation rules for the spin-momentum entanglement of two
qubits[11]. Recent work of Alsing and Milburn extended the results
to situations where one observer is accelerated[8]. In this paper,
we discuss a possible extension to the gravitation field of the
quantum teleportation. Different from the standard teleportation
protocol, our scheme proposes that the observer Alice is
stationary and stays at a place far from a Schwarzschild black
hole, where can be regarded as an asymptotically flat region, and
the other observer Bob passes near her. They coincide at a point
in the flat region, where they instantaneously share an entangled
Bell state. Then Bob picks his qubit travelling to the surface of
a Schwarzschild black hole. In order to ensure Bob's trajectory
does not influence his description of the qubit state, Bob can
access the black hole by free falling. Before he crosses the
horizon, he should accelerate and stop outside the horizon. Thus,
teleportation can be performed between the two observers.\\
\begin{figure}
      \psfig{file=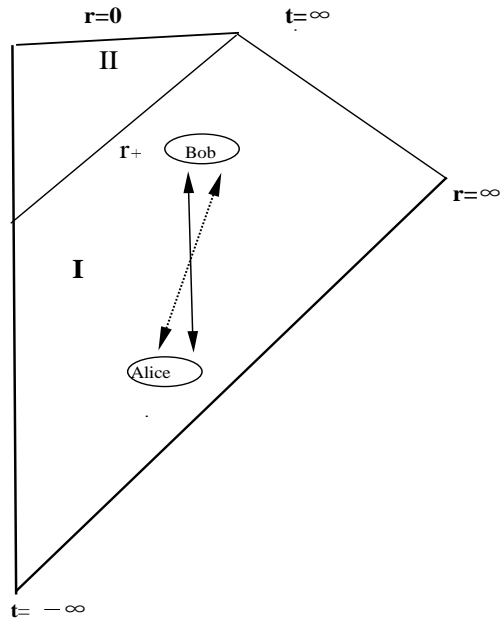,height=8in,width=6in}\caption{
      Penrose diagram
for quantum teleportation in the background of Schwarzschild
space-time. Alice is stationary in the asymptotically flat region,
while Bob locates near the event horizon. With Alice sending the
result of her measurement to Bob, say by photons, Bob will
eventually receive them, and be able to rotate his half of the
shared entangled state into the state $ \mid
\varphi>_{I}=\alpha\mid 0>_{I} +\beta\mid 1>_{I }$}
\end{figure}
\hspace*{7.5mm}Before we discuss the nature of entanglement, we
wish to review the essential features of quantum field theory in
     Schwarzschild space-time[13-16]. It is well known that the Minkowski vacuum state and the vacuum of
a curved space-time are inequivalent. In Rindler frames, it was
found that an accelerated detector becomes spontaneously excited
while moving through the Minkowski vacuum, a result called Unruh
effect. In the area near the event horizon of a Schwarzschild
black hole, the vacuum is populated by particles and antipartiles
as the results of Hawking effect. In the following, we use
two-photon states of the electromagnetic field, which are modelled
by the massless modes of a scalar field, and ignore the
polarization[8,9]. The metric for Schwarzschild space-time is
given by (Plank units are used $\hbar=G=c=k=1$
hereafter)\begin{equation}
ds^2=(1-\frac{2M}{r})dt^2-(1-\frac{2M}{r})^{-1}dr^2-r^2(d\theta
^2+sin\theta^2 d\varphi^2),\end{equation} where the event horizon
is given by $r_{+}=2M$. A complex massless scalar quantum field
$\phi(x)$ in the D-dimensional Minkowski space
     -time can be decomposed in Minkowski modes $\{U_{k}(x)\}$,
     which goes as
     \begin{equation}\phi(x)=\sum_{k}\left[a_{k}U_{k}(x)+{a}_{k}^{\dagger}U_{k}^{\ast}(x)\right],\end{equation}
     where ${a}_{k}$, ${a}_{k}^{\dagger}$ are annihilation and creation operators
     respectively,
     the boundary conditions $k_{n}^{i}=2\pi L_{i}^{-1}n_{i}$ ($i=1,...,D$) and $k=(k_{1},\vec{k})$.
     The Minkowski vacuum can be defined by
     \begin{equation}a_{k}\mid 0_{M}>=0, \forall  k.\end{equation}
By solving the Klein-Gordon equation in the coordinates of
Schwarzschild space-time, the field $\phi(x)$ can be expanded in
normal modes:
\begin{equation}\phi(x)=\sum_{\sigma}\sum_{p}\left[b_{p}^{(\sigma)}u_{p}^{(\sigma)}(x)
+{b}_{p}^{(\sigma)\dag}u_{p}^{(\sigma)*}(x)\right],\end{equation}
 where the operators $b_{p}^{(\sigma)}$ and ${b}_{p}^{(\sigma)\dagger}$
 are assumed to satisfy the usual canonical communication
 relations, $p=(\Omega,\vec{k})$, and the symbol $\sigma=\pm$ refers to region I and II respectively, which is
 separated by the event horizon. By introducing the Unruh modes[15]
 \begin{equation}d_{p}^{(\sigma)}=\int_{-\infty}^{\infty} d\vec{k} p_{\Omega}^{(\sigma)}
 (\vec{k})a_{k_{1},\vec{k}},\end{equation}
  where $\{p_{\Omega}^{(\sigma)}(\vec{k})\}$
 is the complete set of orthogonal functions, the Schwarzschild modes $b_{p}^{(\sigma)}$ and ${b}^{(-\sigma)}_{\tilde{p}}$
 can be well expressed in terms of the Unruh modes by the Bogolubov
 transformations[14]
 \begin{eqnarray}
 b^{(\sigma)}_{p}=[2sinh(4\pi M\Omega)]^{-\frac{1}{2}}\left[e^{2\pi
M \Omega}d^{(\sigma)}_{p}+e^{-2\pi M
 \Omega}{d}^{(-\sigma)\dagger}_{ \tilde{p}}\right],\\
 {b}^{(-\sigma)}_{\tilde{p}}=[2sinh(4\pi M \Omega)]^{-\frac{1}{2}}\left[e^{-2\pi
M  \Omega}d^{(\sigma)\dagger}_{ p}+e^{2\pi M
 \Omega}{d}^{(-\sigma)}_{\tilde{ p}}\right],\end{eqnarray}
 where $\tilde{p}=(\Omega,-\vec{k})$. One then obtains[14]
\begin{equation}
\mid0>_{M}=Z \prod_{\sigma,p}exp(tanhr
b_{p}^{(\sigma)\dagger}{b}_{\tilde{p}}^{(-\sigma)\dagger})\mid0^{(+)}>_{I}\otimes
|0^{(-)}>_{II},\end{equation} where Z is a normalization constant
$Z=\prod_{p}cosh^{-2}r$, where $r=r(p)$,
 $tanhr=e^{-2\pi
M\Omega}$, and $coshr=(1-e^{-4\pi M\Omega})^{-1/2}$. Eq.(8) can be
rewritten as
\begin{equation}
\mid 0>_{M}=Z\prod_{p}(1+...+\frac{1}{n!}tanh^{n}r
b_{p}^{n(\sigma)\dagger}{b}_{\tilde{p}}^{n(-\sigma)\dagger}+...)\mid0>_{I}\otimes
|0>_{II}.\end{equation} Assume \begin{eqnarray}
T^{(+)}(r)=-\sum_{p}(b_{p}^{(+)\dagger}b_{p}^{(+)}lnsinh^{2}r-b_{p}^{(+)}b_{p}^{(+)\dagger}lncosh^{2}r
),\nonumber\\
T^{(-)}(r)=-\sum_{p}(b_{\tilde{p}}^{(-)\dagger}b_{\tilde{p}}^{(-)}lnsinh^{2}r-b_{\tilde{p}}^{(-)}b_{\tilde{p}}^{(-)\dagger}lncosh^{2}r
).\end{eqnarray} The following can be easily proved
\begin{eqnarray}
e^{-T^{(\sigma)}(r)/2}b_{p}^{(\sigma)\dagger}e^{-T^{(\sigma)}(r)/2}=
b_{(p)}^{(\sigma)
\dagger}tanhr,\nonumber\\
e^{-T^{(\sigma)}(r)/2}|0>_{I,II}=\prod_{p}cosh^{-2}r\mid0>_{I}\otimes
|0>_{II}.\end{eqnarray}  By using Eqs.(29), one can rewrite
Eq.(27) as
\begin{eqnarray}
|0>_{M}&=&\left(e^{-T^{(\sigma)}/2}e^{\sum_{p}b_{p}^{(\sigma)
\dagger}b_{\tilde{p}}^{(-\sigma)}}e^{T{(\sigma)}/2}\right)\left(e^{-T^{(\sigma)}/2}|0>_{I,II}\right)\nonumber\\
&=&e^{-T^{(\sigma)}/2}\sum_{n_{p}=0}^{\infty}|n_{p}>_{I}\otimes
|n_{p}>_{II}\nonumber\\
&=&e^{\sum_{p}\left[n_{p}lnsinhr-(1+n_{p})lncoshr\right]}
\sum_{n_{p}=0}^{\infty}|n_{p}>_{I}\otimes|n_{p}>_{II}\nonumber\\
&=&\sum_{n_{p}=0}^{\infty}\prod_{p}tanh^{n_{p}}rcosh^{-1}r|n_{p}>_{I}\otimes
 |n_{p}>_{II},
\end{eqnarray}where $N_{p}=b_{p}^{(\sigma)\dagger}b_{p}^{(\sigma)}$, $b_{p}^{(\sigma)}b_{p}^{(\sigma)\dagger}=1+N_{p}$.
 From Eq.(5),
we see that a given Minkowski mode of frequency $\omega_{\vec{k}}$
is spread over all positive Schwarzschild frequencies $\Omega$, as
a result of the Fourier transform relationship between
$a_{k_{1},\vec{k}}$ and $d^{(\sigma)}_{p}$. One can simplify the
analysis by considering the effect of teleportation of the state
 $\mid \varphi>_{M} = a\mid 0>_{M}+b \mid1>_{M} $ by the Minkowski observer Alice to a single Schwarzschild mode of the
 observer Bob. Thus, one can only consider the mode $p$ in region I which is distinct
 from the mode $\tilde{p}$ in the same region[8]. From Eq.(8), one can find
 that the Minkowski vacuum appears as an entangled state of Hawking particles with nonlocal
 Einstein-Podolsky-Rosen type correlations, that is to say
 \begin{eqnarray}
 \mid 0>_{M}&&=Z\left[\mid 0>_{I}\otimes\mid 0>_{II}+\sum_{p}tanhr(\mid 1_{p}>_{I}^{(+)}
 \otimes\mid 1_{\tilde{p}}>_{II}^{(-)}
+\mid 1_{\tilde{p}}>_{I}^{(+)}\right.\nonumber\\
&&\left.\otimes\mid 1_{p}>_{II}^{(-)}+...+\mid n_{p}>_{I}^{(+)}
 \otimes\mid n_{\tilde{p}}>_{II}^{(-)}
\right.\nonumber\\
&&\left.+\mid n_{\tilde{p}}>_{I}^{(+)})\otimes\mid
1_{p}>_{II}^{(-)}+...\right].\end{eqnarray} Thus we select the
terms with $\sigma=+$ in Eq.(13), which corresponds to the
correlated mode pair in the first set of parentheses, and drop the
unessential phase factors $\{p_{\Omega}^{(\sigma)}(\vec{k})\}$.
Therefore, the single mode component of the Minkowski vacuum
state, namely the two-mode squeezed states, is given by
\begin{equation}
\mid 0>_{M}=\frac{1}{cosh r}\sum^{\infty}_{n=0}
tanh^{n}r|n>_{I}\otimes|n>_{II},
\end{equation}
and similarly \begin{equation} \mid 1>_{M}=\frac{1}{cosh ^2
r}\sum^{\infty}_{n=0}
tanh^{n}r\sqrt{n+1}|n>_{I}\otimes|n>_{II}.\end{equation}
\hspace*{7.5mm}We suppose that Alice and Bob each hold an optical
cavity and they coincide at a point when Bob's frame is
instantaneously at rest. Then, Alice situates in a asymptotically
flat region and Bob travels to the event horizon. Since the
details of Bob's trajectory may influence the Bob's description of
state, for instance, Bob's cavity might be teemed with the thermal
Hawking flux coming out of the black hole. Fortunately, if Bob
access the black hole by free falling, this situation can be
avoided. Near the horizon, the Schwarzschild metric can be
approximately written as
\begin{equation}
ds^2=\frac{r-2M}{2M}dt^2-\frac{2M}{r-2M}dr^2-(2M)^2
(d\theta^2+sin^2\theta d\phi^2),\end{equation} which can be
transformed into cylindrical metric
\begin{equation}
ds^2=d\tau^2-d\rho^2-(2M)^2(d\theta^2+sin^2\theta
d\phi^2).\end{equation} The geodesic near the horizon are
essentially the geodesics of this cylindrical metric. A geodesic
detector/observer will therefore see no particle coming out of the
black hole[14]. Thus, after the two observer's coincidence, Bob
can fall freely into the black hole and manage to stop on the
surface of the black hole by accelerating. In fact, at any fixed
position outside the horizon, the static observer must accelerate
to stay in place and experience a thermal flux of Hawking
radiation, which is analogous to the Unruh radiation in Minskowski
space[17]. We will show in the following how the thermal Hawking
flux of particles influence the fidelity of teleportation.\\
 \hspace*{7.5mm}One can assume
that prior to their coincidence, Alice and Bob have no photons in
their cavities. Suppose that each cavity supports two orthogonal
modes, with the same frequency, labelled $A_{i}, R_{i}$ with
$i=1,2$, which are each excited to a single photon Fock state at
the coincidence point. The state held by Alice and Bob is then the
entangled Bell state
\begin{equation} \mid \varphi>_{M}=\frac{1}{\sqrt{2}}\left(\mid
0>_{M}\mid
    0>_{M}
    +\mid1>_{M}\mid1>_{M}\right),\end{equation} where the first qubit in each term
     refers to cavity Alice, the second qubit cavity Bob. The states $\mid 0>_{M}$, $\mid 1>_{M}$ are defined
     in terms of the physical Fock states for Alice's cavity by the dual rail basis states as
     suggested by Ref[9]: $\mid 0>_{M}=\mid 1>_{A_{1}}\mid 0>_{A_{2}}$, $\mid 1>_{M}=\mid
     0>_{A_{1}}\mid1>_{A_{2}}$, and the similar expressions for
     Bob's cavity. In order to teleport the unknown state $\mid \varphi>_{M} = \alpha\mid 0>_{M}+\beta \mid1>_{M} $
     to Bob, we should assume that Alice has an additional cavity,
     which contains a single qubit with dual rail encoding by a
     photon excitation of a two mode Minkowski vacuum state. This
     will allow Alice to perform a joint measurement on the two
     orthogonal modes of each cavity. After Alice's measurements,
     Bob's photon will have been projected according the
     measurements outcome. The final state Bob received can be given by
      $\mid\phi_{ij}>=x_{ij}\mid0>+y_{ij}\mid1>$, where there
      are four possible conditional state amplitudes as
      $(x_{00},y_{00})=(\alpha,\beta),
      (x_{01},y_{01})=(\beta,\alpha),(x_{10},y_{10})=(\alpha,-\beta)$, and
      $(x_{11},y_{11})=(-\beta,\alpha)$. Once receiving Alice's results of
     measurement, Bob can apply a unitary transformation to verify
     the protocol in his local frame. However, Bob must confront
     the fact his cavity will become teemed with thermally excited
     photons because of Hawking effect.\\
\hspace*{7.5mm} When Alice sends
\begin{figure}
\psfig{file=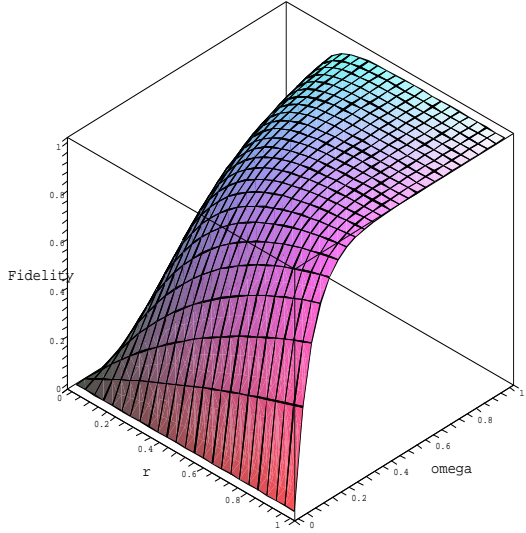,height=3in,width=3in}\caption{The fidelity of
teleportation between Alice and Bob is closely related to the
radius of the black hole. The larger the radius is, the higher the
fidelity can be reached. This figure shows the fidelity of
 teleportation for a small black hole:  $0.0001\leq r
\leq1 (m)$ and $0.001\leq \Omega \leq 1$ (Hz).}
\end{figure}
 the result of her measurement to
Bob, Bob' state will be projected into [8]
\begin{eqnarray}
 \rho^{(I)}_{ij}&&=\sum^{\infty}_{k=0}\sum^{\infty}_{l=0}{}_{II}<k,l\mid\phi_{ij}><\phi_{ij}\mid
 k,l>_{II}\nonumber\\
 &&=\frac{1}{cosh^{r}}\sum^{\infty}_{k=0}\sum^{n}_{l=0}\left[(tanh^{2}r)^{n-1}[(n-m)|x_{ij}|^{2}+m
 |y_{ij}|^{2}]\right.\nonumber\\
 &&\left.\times|m,n-m>_{out}<m,n-m|+(x_{ij}y^{*}_{ij}tanh^{2n}r\sqrt{(m+1)(n-m+1)})\right
 . \nonumber\\
 &&\left .\times |m,n-m+1>_{out}<m+1,n-m|+H.c.)\right]\end{eqnarray}
 Bob's premeasurement state can be written as[8]
 \begin{equation}
 \rho^{I}_{ij}=\sum^{\infty}_{n=0}p_{n}\rho^{I}_{ij,n}\end{equation}
 in particular with
 \begin{equation}
 \rho^{I}_{ij,1}=|\phi_{ij}><\phi_{ij}|,
  p_{0}=0, p_{1}=1/cosh^{6}r.\end{equation}
  Since what we concern about is to which extent $\mid \varphi_{ij}>$
  might deviate from unitarity, it is reasonable for us to
  perform a unitary transformation on $\mid\varphi_{ij}>$ and convert its form into
  $\mid \varphi>$. In this way, we may define the fidelity  corresponds to
  the teleportation, which goes as
\begin{equation}
  F^{I}(\mid \varphi>)={}_{I}<\varphi\mid\rho^{I}\hat{U}\mid\varphi_{ij}>_{I}
  ={}_{I}<\varphi\mid\rho^{I}\mid\varphi>_{I}=1/(cosh^6 r).
\end{equation} The fidelity of teleportation between Alice and Bob can be given
  by $F^{I}=1/cosh^6 r$ which is identical with the celebrated results of
  Alsing and Milburn.  From figure 2, we can see that, if $\Omega$ is fixed, then the fidelity of teleportation
  is closely related to the radius (or mass) of the black hole.
  For a small black hole with the radius of several meters, the
  fidelity is almost unitary. Thus, we can reasonably deduce that
  for a Sun-like black hole (with the mass $M_{\odot}$) the fidelity of
   teleportation is incredibly high.\\
\hspace*{7.5mm}We have demonstrated that the fidelity of
teleportation is reduced seen by Bob who locates near the horizon
of the black hole because of Hawking effect. For teleportation
outside a tiny black hole, all information is lost and what Bob
perceives is only thermalize state. But one can be free from worry
about fidelity reduction when teleportation is conducted outside a
galaxy-like black hole since the fidelity there is high.
\begin{center}\textbf{ACKNOWLEDGEMENTS}\end{center}
The authors would like to thank  the anonymous referee for his
helpful comments and the Interdisciplinary Center for Theoretical
Study at the University of Science and Technology of China, where
part of this work is performed. The work has been supported by the
National Natural Science Foundation of China under Grant No.
10273017.

\end{document}